\documentclass{aa}
\usepackage[varg]{txfonts}

\usepackage{graphicx}
\usepackage{dcolumn}
\usepackage{bm}

\usepackage{amsmath}
\usepackage{amssymb}

\usepackage{euscript}

\newcommand{\cvir}{c_{vir}}
\newcommand{\rvir}{R_{vir}}

\newcommand{\mtot}{M_{tot}}

\begin{document}

\titlerunning{Why does Einasto profile index $n\sim 6$ occur so frequently?}
\title{Why does Einasto profile index $n\sim 6$ occur so frequently?}

\author{Anton N. Baushev\inst{1, 2, 3}\and Maxim V. Barkov\inst{1,4,5}}
\institute{DESY, 15738 Zeuthen, Germany\and
 Institut f\"ur Physik und Astronomie, Universit\"at Potsdam, 14476
Potsdam-Golm, Germany\and Bogoliubov Laboratory of Theoretical Physics, Joint Institute for Nuclear
Research, 141980 Dubna, Moscow Region, Russia \and Department of Physics and Astronomy, Purdue University,
525 Northwestern Avenue, West Lafayette, IN 47907-2036, USA \and Astrophysical Big Bang Laboratory, RIKEN, 2-1
Hirosawa, Wako, 351-0198 Saitama, Japan }
 \offprints{baushev@gmail.com}

\date{}

\abstract{We consider the behavior of spherically symmetric Einasto halos composed of gravitating
particles in the Fokker-Planck approximation. This approach allows us to consider the undesirable
influence of close encounters in the N-body simulations more adequately than the generally accepted
criteria. The Einasto profile with index $n \approx 6$ is a stationary solution of the
Fokker-Planck equation in the halo center. There are some reasons to believe that the solution is
an attractor. Then the Fokker-Planck diffusion tends to transform a density profile to the
equilibrium one with the Einasto index $n \approx 6$. We suggest this effect as a possible reason
why the Einasto index $n \approx 6$ occurs so frequently in the interpretation of N-body simulation
results. The results obtained cast doubt on generally accepted criteria of N-body simulation
convergence.}


\maketitle

\section{Introduction}
The Einasto density profile was first suggested in 1965 \citep{einasto} and since then has been
extensively used for fitting stellar systems and results of N-body simulations. The profile has the
shape of $\rho(r)=\rho_c \exp\left[-2 n\left\{ (r/r_s)^{1/n}\right\}\right]$ where $\rho_c$ is the
central density, $r_s$ is the characteristic radius, and $n$ is the profile index\footnote{This
definition is equivalent to the standard one, if we introduce $\rho_s=\rho_c\exp\left(-2 n\right)$
and $\rho=\rho_s \exp\left[-2 n\left\{ (r/r_s)^{1/n}-1\right\}\right]$.}. The lower $n$ is, the
shallower the central peak is: $n\lesssim 1$ corresponds to cored profiles, $n>1$ --- to the cuspy
ones; $\alpha\equiv 1/n$ is often used instead of $n$.

We will mainly focus on the Einasto profile application in the treatment of N-body simulation
results. The profile fits recent simulations even better than the Navarro-Frenk-White (NFW) one;
the profile index turns out to be quite universal $n\simeq (6-7)$ ($\alpha\simeq (0.15-0.17)$)
\citep{gao2008, navarro2010, dutton2014}. Since the index is quite high, the profile suggested by
the N-body modeling is cuspy.

On the contrary, observations suggest 'cored' profiles for, at least, dark matter (DM) halos of
dwarf galaxies (see, for instance, \citep{mamon2011, walker2011}). In particular, the Einasto
profile was used to fit observational data for a large data set of various galaxies
\citep{mamon2011}. The Einasto index $n\simeq 0.5$ was obtained, which corresponds to a cored
profile. The contradiction is well-known as the 'core-cusp problem' and might indicate that the
cold noninteracting DM paradigm was inadequate. However, some evidences from theoretical
consideration \citep{13} and simulations \citep{17} have been reported that the cusps appearing in
the N-body simulations might be just a numerical artifact. We will discuss this point at great
length in the two last sections, here restricting ourselves to a short introduction to the problem.

As a rule, the Einasto profile is used to fit systems that may be considered to a first
approximation as collisionless. However, the collisions are not always absent completely. Stellar
systems and N-body models are constructed by point-mass particles with a newtonian or
quasi-newtonian gravitational field. Therefore, they may scatter on the individual potential wells
surrounding each particle. On the contrary, real dark matter systems should be completely
collisionless, as we shall show below.

The encounters may be classified into close and weak, depending on the ratio between the actual
impact parameter $b$ and the $90$ degree deflection impact parameter $b_{90}$. Each close one leads
to a strong change of momenta of the colliding particles. However, they are quite rare in stellar
systems ($b_{90}$ is typically very small). N-body simulations are freed of them by the newtonian
potential smoothing. We will not consider close encounters in this paper.

On the contrary, each weak encounter changes the particle momenta only slightly. The combined
effect of many weak collisions is a slow diffusion of particles in the phase space. Since the
gravitational force is long-acting, the weak encounter effect dominates. Its importance can be
roughly estimated with the help of collisional relaxation time \citep[chapter 1.2.1]{bt}
\begin{equation}
\tau_r= \dfrac{N(r)}{8 \ln\Lambda} \tau_d.
 \label{relaxation_time}
\end{equation}
Here $N(r)$ is the number of particles inside the radius $r$,  $\tau_d=(4\pi
G\bar\rho(r)/3)^{-1/2}$ is the characteristic dynamical time of the system at the radius $r$,
$\bar\rho(r)$ is the average density inside $r$, $\Lambda=b_{max}/b_{min}$ (where $b_{max}$ and
$b_{min}$ are the characteristic maximum and minimum values of the impact parameter) is the Coulomb
logarithm; $b_{max}$ is typically comparable with the size of the halo; $b_{min}$ is the radius
where either the assumption of a straight-line trajectory breaks, or the newtonian potential is no
longer valid. Since $\ln\Lambda$ depends on $b_{max}$ and $b_{min}$ only logarithmically, it is
usually enough to perform a rough estimation of these quantities. For stellar systems
$\Lambda\sim\rvir/b_{90}\simeq \rvir v^2/(Gm)\simeq N$. For N-body simulations $\Lambda\sim
r_s/\varrho$, where $\varrho$ is the potential softening radius: for instance, $\Lambda= 3
r_s/\varrho$ \citep{klypin2013} or $\Lambda= \min[N, r_s/(4 \varrho)]$ \citep{farouki1982}.

For stellar systems, the weak encounters are quite physical. On the contrary, the encounters are no
more than an undesirable numerical effect in the case of N-body simulations. Indeed, a real DM halo
contains some $\sim 10^{60}$ particles, and its collisional relaxation time (\ref{relaxation_time})
is gigantic. The number of test bodies in N-body simulations is $\sim 50$ orders of magnitude
lower. Therefore, their collisions may produce a dynamical friction and other unphysical relaxation
processes. One needs to be secured in the negligibility of the collisions to guarantee the
reliability of N-body simulations. To put it differently, the ratio $t_0/\tau_r$, where $t_0$ is
the lifetime of the system, should be low enough. An estimation of the maximum ratio $t_0/\tau_r$
wherein N-body simulation results are still not corrupted by the collisions is one of the main aims
of N-body convergency tests. The commonly used criteria of N-body simulation convergency are
entirely based on merely the density profile stability (for instance, \citep{power2003}).

In this paper, we study the influence of the weak encounters in the case of the Einasto profile
with the help of the Fokker-Planck (hereafter FP) equation. The results of the study can be
interesting for stellar system dynamics. Moreover, with the help of the FP approach, we may
consider the behavior of N-body simulation particles: since the strong encounters are totally
excluded here by the potential softening, the FP approximation perfectly works in this case. In
section 2, we set forth the mathematical formalism and the Fokker-Planck equations for the Einasto
profile. In section 3, we discuss the results obtained and suggest a possible reason why the value
$n\simeq (6-7)$ may be distinguished and is so widely occurring in the simulations.

\section{Calculations: basic equations}
We assume that the system under consideration consists of identical particles of mass $m$. It can
be described by the particle distribution function $k$.

We define the gravitational potential $\phi$ as the minus potential energy of a unit mass. The
potential at infinity is chosen to be zero. Thus, $\phi$ is everywhere positive and reaches its
maximum in the halo center. We find it convenient to define the particle specific energy
${\EuScript E}$ in the gravitational field $\phi$ as ${\EuScript E}=\phi-v^2/2$, where $v$ is the
particle velocity. ${\EuScript E}$ is always positive for a bound particle; $k=0$ if ${\EuScript
E}\le 0$. For brevity sake, the specific energy will be referred to as "energy." Since we will
never use the real particle energy $-m {\EuScript E}$ or potential energy $-m \phi$, this
definition cannot lead to misunderstanding.

The halo mass inside the radius $r$ is $M(r)=\int^r_0 4\pi r^2 \rho(r) dr$. It is convenient to
introduce a dimensionless radius $y=r/r_s$. We obtain
\begin{eqnarray}
M(r)=\mtot\left(1-\frac{\Gamma(3n, 2n y^{1/n})}{\Gamma(3n)}\right);\quad \mtot=\dfrac{4\pi \rho_c
r^3_s n \Gamma(3n)}{(2n)^{3n}},\nonumber\\
N(r)=N_{tot}\left(1-\frac{\Gamma(3n, 2n y^{1/n})}{\Gamma(3n)}\right);\quad N_{tot}=\dfrac{4\pi
\rho_c r^3_s n \Gamma(3n)}{m (2n)^{3n}},\nonumber\\ \label{18a5}
\end{eqnarray}
where $\mtot$ and $N_{tot}$ are the total mass and the number of particles in the halo. $\Gamma$ is
the gamma function
\begin{eqnarray}
 \label{18a6}
&\Gamma(n, y)=\int^\infty_y x^{n-1}e^{-x} dx;\quad \gamma(n, y)=\int^y_0 x^{n-1}e^{-x}
dx\\
&\Gamma(n)=\Gamma(n, 0)=\int^\infty_0 x^{n-1}e^{-x} dx\qquad \Gamma(n, y)+\gamma(n,
y)=\Gamma(n)\nonumber
\end{eqnarray}
We denote the value of the halo center potential by $W=\dfrac{G \mtot}{r_s} \dfrac{(2n)^n
\Gamma(2n)}{\Gamma(3n)}$. Then the potential can be written as
\begin{equation}
\phi(r)=\frac{W}{\Gamma(2n)}\left(\frac{\gamma(3n, 2n\cdot y^{1/n})}{(2n)^n y}+ \Gamma(2n, 2n\cdot
y^{1/n})\right)
 \label{18b1}
\end{equation}

In this paper, we perform an accurate consideration of the process with the help of the
Fokker-Planck equation \citep{ll10}. We are interested in the central region of the halo ($r<r_s$),
since the density here is the highest, and so the encounters are the most pronounced. We exactly
follow the method used by \citep{evans1997} to apply the FP approach in the halo center. They
supposed that the system is spherically symmetric and isotropic. The latter supposition is quite
natural for the halo center: the velocity distribution becomes isotropic in this region even if it
is extremely anisotropic at larger radii \citep{14}. For a spherically symmetric, isotropic system,
the particle distribution $k$ depends only on the energy $\EuScript E$ and time \citep{bt} (i.e.
the number of particles with energies between $\EuScript E$ and ${\EuScript E}+d{\EuScript E}$ is
$k({\EuScript E},t) d{\EuScript E}$), and the FP equation reads
\begin{equation}
\frac{\partial k({\EuScript E},t)}{\partial t}=\frac{\partial }{\partial {\EuScript E}}\left\{
k({\EuScript E},t) D_1({\EuScript E}) +\frac{\partial}{\partial {\EuScript E}}[k({\EuScript E},t)
D_2({\EuScript E})/2]\right\},
 \label{18a3}
\end{equation}
where $D_1({\EuScript E})$ and $D_2({\EuScript E})$ are the diffusion coefficients. The encounters
result in a slow diffusion of particles in the phase space. The diffusion flux $s$ of particles is
\begin{equation}
s({\EuScript E})= k({\EuScript E},t) D_1({\EuScript E}) +\frac{\partial}{\partial {\EuScript
E}}[k({\EuScript E},t) D_2({\EuScript E})/2],
 \label{18a4}
\end{equation}
and the FP equation can be rewritten in the form of the particle number conservation $\partial
k({\EuScript E},t)/\partial t=-\mathop{\rm div} \vec s$.  In the case of one-dimensional task under
consideration, $s$ is just the number of particles crossing the surface ${\EuScript E}=\text{\it
const}$ per unit time.

Now we should define the characteristic time $\tau_{FP}(r)$ in which the Fokker-Planck diffusion
changes the Einasto profile significantly. The maximal reasonable estimation of $\tau_{FP}$ is
\begin{equation}
 \label{18e1}
 \tau_{FP,max}(r)=\frac{N(r)}{dN/dt}=\frac{N(r)}{s(r)}
\end{equation}
Indeed, this value corresponds to the time sufficient for the FP diffusion to remove all the
particles inside $r$. However, $\tau_{FP,max}$ gives a strongly overestimated value of
$\tau_{FP}(r)$: the number of particles sufficient to change the profile inside this radius is
significantly smaller. As an example, let us consider the fraction of mass (or, which is the same,
of the test particles) one should add to the $\rho\propto r^{-1}$ profile in order to transform it
into $\rho\propto r^{-4/3}$ (the importance of this particular instance will be clear from
section~\ref{results}). To be more precise, let us consider a $\rho\propto r^{-1}$ profile that has
density $\rho_0$ at some radius $l$. Then the profile density is $\rho=\rho_0 (r/l)^{-1}$, and the
mass inside $l$ is $m_l=2\pi l^3 \rho_0$. What fraction of $m_l$ should be added to the profile in
order to transform it into $\rho\propto r^{-4/3}$ so that the density at $l$ remains the same? The
profile after the modification should be $\rho=\rho_0 (r/l)^{-4/3}$, and a trivial integration
shows that its mass inside $l$ is $1.2 m_l$, i.e., only $20\%$ of the initial mass should be added.

In a like manner, for each radius $r$ we can determine the number of particles $N_{+1}(n, r)$ that
should be added to the Einasto profile with index $n$ in order to transform it inside $r$ into the
Einasto profile with index $n+1$, but with the same density at $r$ and the same value of $r_s$. The
final profile has the index $n+1$, and its central density $\rho_{c,f}$ is defined by the
requirement of the equal density at $r$: $\rho_{c,f} \exp\left[-2 (n+1)\left\{
(r/r_s)^{1/(n+1)}\right\}\right]=\rho_c \exp\left[-2 n\left\{ (r/r_s)^{1/n}\right\}\right]$. Then
equation~(\ref{18a5}) allows us to calculate $N_{+1}$ as a function of $n$ and $r$. It is natural
to define $\tau_{FP}(r)$ as
\begin{equation}
 \label{18a42}
 \tau_{FP}(r)=\frac{N_{+1}(r)}{s(r)}
\end{equation}
Being so defined, $\tau_{FP}(r)$ characterizes the time it takes for the FP diffusion $s(r)$ to
bring through the radius $r$ a number of particles sufficient to change the profile index inside
this radius by one.

To show the direction of the FP diffusion, we introduce the sign of $\tau_{FP}(r)$ and
$\tau_{FP,max}(r)$: it coincides with that of $s(r)$. The density profile is mainly determined by
the particle collisions if $t\ge \tau_{FP}$. For the important instance of N-body simulations, it
means that they can nohow model a real collisionless DM system unless $t_0$ is significantly lower
than $\tau_{FP}$.

Equation~(\ref{relaxation_time}) estimates the relaxation time rather crudely. A much more reliable
estimation can be derived from the Fokker-Planck equation \citep[eqn. 7.106]{bt}, which yields that
\begin{equation}
 \label{18e2}
 \tau_{r,2}= 0.34 \dfrac{\sigma^3}{G^2 m \rho \ln\Lambda}.
\end{equation}
Here $\sigma$ is the local one-dimensional velocity dispersion, $\rho$ is the local density, and
$G$ and $m$ are the gravitational constant and the particle mass, respectively. We will use both
definitions of the relaxation time, $\tau_r$~(\ref{relaxation_time}) and $\tau_{r,2}$~(\ref{18e2}).
The second one is much more precise, but the first one is widely used in literature (for instance,
by \citep{power2003}).

The diffusion coefficients read \citep{evans1997}
\begin{eqnarray}
D_1({\EuScript E})=\frac{16\pi^2 G^2 m^2 \ln\Lambda}{p({\EuScript E})}\left[p({\EuScript
E})\!\!\int\limits^{\EuScript E}_0\!\! f(\grave {\EuScript E}) d\grave {\EuScript
E}-\int\limits^W_{\EuScript E}\!\! f(\grave
{\EuScript E}) p(\grave {\EuScript E}) d\grave {\EuScript E}\right],\label{18a8}\\
D_2({\EuScript E})=\frac{32\pi^2 G^2 m^2 \ln\Lambda}{p({\EuScript
E})}\left[\int\limits^W_{\EuScript E}\!\! q(\grave {\EuScript E}) f(\grave {\EuScript E}) d\grave
{\EuScript E}+q({\EuScript E})\!\!\int\limits^{\EuScript E}_0\!\! f(\grave {\EuScript E}) d\grave
{\EuScript E}\right]. \nonumber
\end{eqnarray}
Here, $p({\EuScript E})$ and $q({\EuScript E})$ are the density of states and the total phase-space
volume with energy between ${\EuScript E}$ and $W$, respectively
\begin{eqnarray}
p({\EuScript E})=16\pi^2\int^{\large r_{max}({\EuScript E})}_0 \left(2[\phi(r)-{\EuScript E}]\right)^{1/2} r^2 dr, \label{18a9}\\
q({\EuScript E})=\frac{16\pi^2}{3}\int^{\large r_{max}({\EuScript E})}_0 \left(2[\phi(r)-{\EuScript
E}]\right)^{3/2} r^2 dr. \label{18a10}
\end{eqnarray}
The maximum radial digression $r_{max}({\EuScript E})$ possible for a particle of energy
${\EuScript E}$ can be obtained from the energy conservation law
\begin{equation}
\phi(r_{max}({\EuScript E}))={\EuScript E} \label{18a11}.
\end{equation}
The phase-space density $f({\EuScript E})$ of particles reads \citep[eqn. 4.46b]{bt}
\begin{equation}
f({\EuScript E})=\frac{1}{\sqrt8\pi^2\mtot}\int^{\EuScript E}_0\!\!\frac{d\phi}{\sqrt{{\EuScript
E}-\phi}}\:\frac{d^2\rho}{d\phi^2}.\label{18a12}
\end{equation}
where ${d^2\rho}/{d\phi^2}$ is the second derivative of the halo density $\rho$ as a function of
the gravitational potential $\phi$. The equations for $\rho(r)$ and (\ref{18b1}) implicitly define
this function. Finally, we need to mention an important relationship between the particle
distribution function $k({\EuScript E},t)$, the phase-space distribution function $f({\EuScript
E},t)$ and the density of states $p({\EuScript E})$:
\begin{equation}
k({\EuScript E},t)=f({\EuScript E},t)\cdot p({\EuScript E}).\label{18a14}
\end{equation}

One can see that $q$, $p$, $s$, and $f$ are defined as functions of $\EuScript E$, while $N$, $M$,
and $\phi$ depend on $r$. In order to close the system of equations (for instance, to calculate
$\tau_{FP}$ in accordance with (\ref{18a42})), we need to bind $\EuScript E$ and $r$ (or $\EuScript
E$ and $y$). Theoretically, there is no one-to-one correspondence between the particle energy and
the maximal radius of its orbit: it depends on the orbit shape. For a purely radial orbit
\begin{equation}
\EuScript E=\frac{W}{\Gamma(2n)}\left(\frac{\gamma(3n, 2n\cdot y^{1/n})}{(2n)^n y}+ \Gamma(2n,
2n\cdot y^{1/n})\right).
 \label{18c13}
\end{equation}
For a circular orbit
\begin{equation}
\EuScript E=\frac{W}{\Gamma(2n)}\left(\frac{\gamma(3n, 2n\cdot y^{1/n})}{2 (2n)^n y}+ \Gamma(2n,
2n\cdot y^{1/n})\right).
 \label{18c14}
\end{equation}
The difference between these equations is not very large. Since we consider the case when particle
velocity distribution is isotropic, the orbits of the bulk of particles are more or less circular.
Therefore, we use relationship (\ref{18c14}) between $\EuScript E$ and $r$.

\section{Calculations: numerical solution}
One can see that equations (\ref{18a5})-(\ref{18a12}) form a closed system. Here we represent the
main course of numerical solution of the equations in the interval $n\in [1;12]$.

First of all, it is convenient to turn to dimensionless quantities. We will use $y\equiv r/r_s$ and
$E\equiv \EuScript E/W$ as coordinates. The dimensionless potential can be introduced as
$\psi\equiv \phi/W$
\begin{equation}
\psi(y)=\frac{1}{\Gamma(2n)}\left(\frac{\gamma(3n, 2n\cdot y^{1/n})}{(2n)^n y}+ \Gamma(2n, 2n\cdot
y^{1/n})\right)
 \label{18c1}
\end{equation}
We can also introduce dimensionless equivalents for $p(E)$, $q(E)$, $f(E)$, $s(E)$, and dynamical
time, denoting them by the same letter with a tilde above:
\begin{align}
p(E)&=16\sqrt{2}\pi^2 r^3_s \sqrt{W}\tilde p(E); \quad q(E)=16\sqrt{2}\pi^2 r^3_s W\sqrt{W}\tilde q(E)\nonumber\\
f(E)&=\frac{\rho_c \Gamma(2n)}{\sqrt{2}\pi^2 W\sqrt{W} (2n)^{2n}} \tilde f(E); \qquad
\tau_d= \frac{r_s}{\sqrt{W}} \widetilde{\tau_d}\label{18d1}\\
\frac{d^2\rho}{d\phi^2}&=\frac{2\rho_c \Gamma(2n)}{W^2 (2n)^{2n}}
\widetilde{\frac{d^2\rho}{d\phi^2}};\qquad s(E)=\frac{8\sqrt{2W} \ln\Lambda}{\pi^2 n^{2} r_s}
\tilde s(E) \nonumber
\end{align}
After some trivial but bulky calculations, we obtain from equations (\ref{18a5})-(\ref{18a12}):
\begin{eqnarray}
\tilde p(E)=\int^{\large y_{max}}_0 y^2 \left(\psi(y)-E\right)^{1/2} dy, \label{18c2}\\
\tilde q(E)=\frac23 \int^{\large y_{max}}_0 y^2 \left(\psi(y)- E\right)^{3/2} dy. \label{18c3}
\end{eqnarray}
where $y_{max}$ is defined by (\ref{18a11})
\begin{equation}
\psi(y_{max})=E \label{18c9}
\end{equation}
\begin{equation}
\tilde f(E)=\int^E_0\!\!\frac{d\psi}{\sqrt{E-\psi}} \: \widetilde{\frac{d^2\rho}{d\phi^2}}
\label{18c4}
\end{equation}
where
\begin{eqnarray}
\widetilde{\frac{d^2\rho}{d\phi^2}}(y)=(2n)^{4n}\Gamma(2n) \dfrac{y^{2+\frac{1}{n}}\exp(-2n\cdot
y^{1/n})}{\gamma^2(3n, 2n\cdot y^{1/n})}\times\nonumber\\
\times\left(2y^{1/n}-\frac{n+1}{n}+\dfrac{y^{3}(2n)^{3n}\exp(-2n\cdot y^{1/n})}{n\gamma(3n, 2n\cdot
y^{1/n})}\right)\label{18c5}
\end{eqnarray}
and
\begin{equation}
\tilde s(E)=\frac{d\widetilde{f}}{dE}
\left(\int^1_E\!\!\widetilde{q}\cdot\widetilde{f}dE+\widetilde{q}
\int^E_0\!\!\widetilde{f}dE\right)-\widetilde{f}\int^1_E\!\!\widetilde{p}\cdot\widetilde{f}dE
 \label{18c6}
\end{equation}
\begin{equation}
\widetilde{\tau_d}=\sqrt{\frac{y^3 (2n)^n \Gamma(2n)}{\gamma(3n, 2n\cdot y^{1/n})}}
 \label{18c11}
\end{equation}

\noindent Equation (\ref{18c4}) for $\tilde f(E)$ can be rewritten as
\begin{equation}
\tilde f(E)=\int\limits^\infty_{y_{max}(E)}\!\!\frac{-(d\psi/dy) dy}{\sqrt{E-\psi(y)}} \:
\widetilde{\frac{d^2\rho}{d\phi^2}}(y), \label{18c8}
\end{equation}
where
\begin{equation}
-\frac{d\psi}{dy}=\frac{\gamma(3n, 2n\cdot y^{1/n})}{(2n)^n \Gamma(2n) y^2} \label{18c10}
\end{equation}
To make the integral calculations faster, one may exclude the singularity at $E=\psi$ in
(\ref{18c8}).
\begin{equation}
\tilde f(E)=2 \sqrt{E}\:
\widetilde{\frac{d^2\rho}{d\phi^2}}(E)+\int^E_0\!\!\dfrac{\left(\widetilde{\frac{d^2\rho}{d\phi^2}}(\psi)
- \widetilde{\frac{d^2\rho}{d\phi^2}}(E)\right)}{\sqrt{E-\psi}} \:d\psi  \label{18c7}
\end{equation}

We used the standard Matlab functions for numerical integration \citep{matint}. The relative
accuracy for the internal integrals (Eq.~(\ref{18c2},\ref{18c3},\ref{18c7})) was set at the level
$10^{-5}$ and for the external integral (Eq.~(\ref{18c6})) --- at $10^{-4}$.

\begin{figure}
 \resizebox{\hsize}{!}{\includegraphics[angle=0]{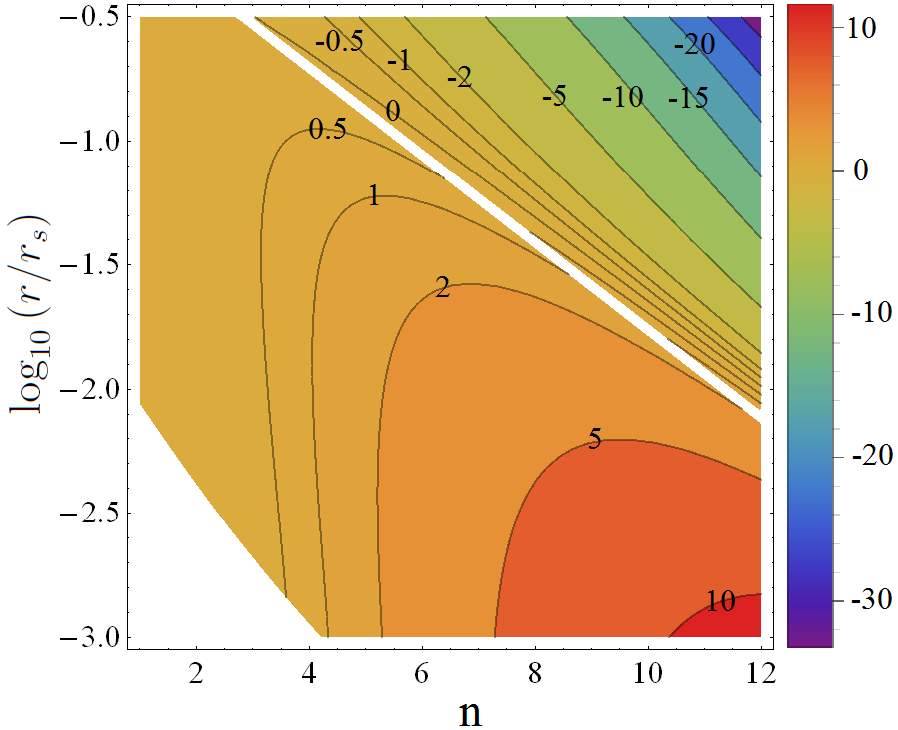}}
\caption{The ratio between the relaxation time $\tau_r(r)$ (defined by~(\ref{relaxation_time})) and
the time in which the Fokker-Planck diffusion can shift the Einasto index $n$ of the profile by
one, $\tau_{FP}(r)$~(\ref{18a42}), as a function of the radius and $n$. The white line represents
the area (\ref{18a13}) where the Einasto profiles behave as the power-law $\rho\propto r^{-4/3}$.
The blank region in the low left corner corresponds to the area uncovered by our calculations.}
 \label{18fig1}
\end{figure}

\section{Results}
\label{results} 
\begin{figure}
\resizebox{\hsize}{!}{\includegraphics[angle=0]{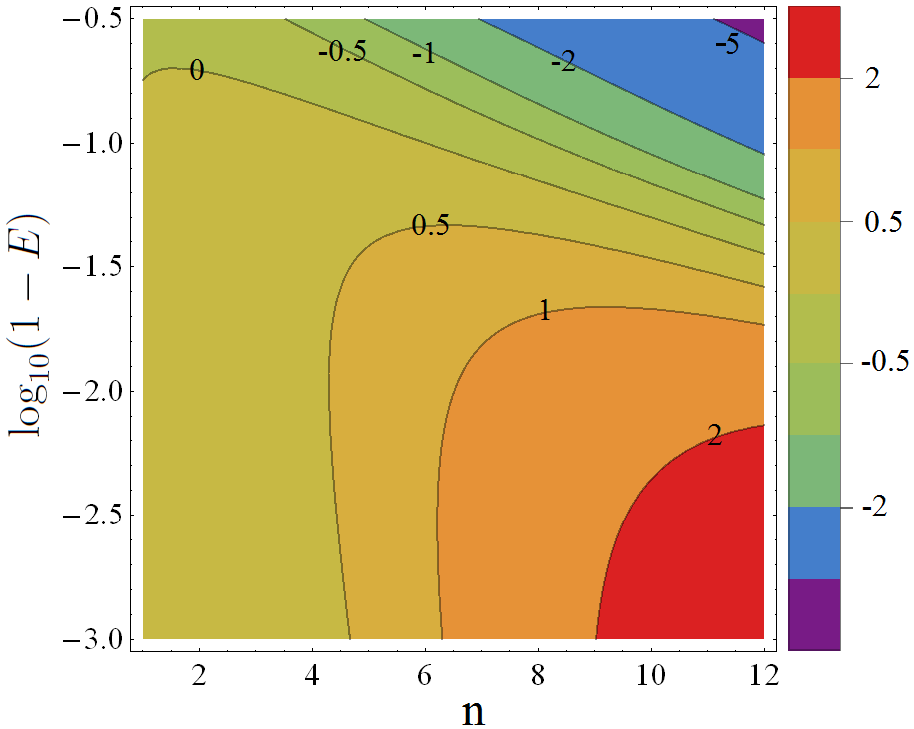}} \caption{The ratio between the
relaxation time $\tau_{r,2}(r)$ (defined by~(\ref{18e2})), and the maximal estimation~(\ref{18e1})
of the Fokker-Planck diffusion time $\tau_{FP,max}(r)$, as a function of dimensionless energy $E$
and $n$.}
 \label{18fig2}
\end{figure}

Figure~\ref{18fig1} represents the ratio between the relaxation time
$\tau_r(r)$~(\ref{relaxation_time}) and the time in which the Fokker-Planck diffusion can shift the
Einasto index $n$ of the profile by one, $\tau_{FP}(r)$~(\ref{18a42}), as a function of the radius
and $n$. The blank region in the low left corner corresponds to the area uncovered by our
calculations. We should remind that the sign of $\tau_r/\tau_{FP}$ coincides with the sign of
$s(r)$ and demonstrates the direction of the FP diffusion. Figure~\ref{18fig2} represents the ratio
between the relaxation time $\tau_{r,2}(r)$~(\ref{18e2}) and the maximal estimation~(\ref{18e1}) of
the Fokker-Planck diffusion time $\tau_{FP,max}(r)$, as a function of dimensionless energy $E$ and
$n$.

The most remarkable feature in both the Figures is the line $\tau_r/\tau_{FP}=0$ and the fact that
FP diffusion changes its direction on this line. The line defines a stationary solution for the
Fokker-Planck equation. Apparently, the Einasto profile cannot have different $n$ at different
radii. However, Fig.~\ref{18fig1} defines the direction of the profile variation caused by the weak
encounters. A similar solution has already been reported in the case of power-law profiles
\citep{evans1997, 13}. The authors of \citep{evans1997} assumed an isotropic velocity distribution
at each point, exactly as we do in this paper. They found that the density profile $\rho\propto
r^{-\gamma}$, where $\gamma=4/3$, is a stationary solution. We can easily compare this result with
ours. The Einasto profile with index $n$ behaves as the power-law with $\gamma=4/3$ at the radius
where $d\log \rho/d\log r= -4/3$. We obtain for the Einasto profile:
\begin{equation}
\log(r/r_s)=n \log\frac23.\label{18a13}
\end{equation}
This equation defines the white straight line in Fig.~\ref{18fig1}. One can see that the line
$\tau_r/\tau_{FP}=0$ almost coincides with it. A small discrepancy occurs as a result of the
coordinate transformation: equations (\ref{18a5})-(\ref{18a12}) define $s$ as a function of
$\EuScript E$. We transform it into the dependence of $r$, supposing that the radius $r$ of the
circular orbit of a particle with energy $\EuScript E$ corresponds to this energy. However, these
dependencies are slightly different for the power-law and Einasto profiles, which accounts for the
deflection of the curve $\tau_r/\tau_{FP}=0$ from (\ref{18a13}).

Thus, the $\tau_r/\tau_{FP}=0$ curve that we found is just the solution $\rho\propto r^{-4/3}$
found by \citep{evans1997}. Two conclusions may be drawn from this fact: first, the results of our
calculations and \citep{evans1997} confirm each other. Second, there is no other stationary
solution of the Fokker-Planck, at least, in the area covered by our simulations. Indeed, the
authors of \citep{evans1997} used power-law profiles to investigate FP diffusion. Since the
stationary solution turns out to be power-law, they discovered it explicitly. We are less fortunate
using the Einasto profiles as test ones. None of them is a stationary solution, and we can see the
stationary profile as a slanting line, but we are still able to find it. Thus, even in the case of
an unfortunate choice of the model profile to investigate the FP equations, the stationary solution
is visible.

A question appears: is the stationary solution $\rho\propto r^{-4/3}$ stable or not? There are some
arguments testifying that the solution is an attractor. First, equation 11 in \citep{evans1997}
shows that the FP diffusion tends to change any power-law profile with an index $\gamma$ close
enough to $4/3$ towards $\rho\propto r^{-4/3}$: if the profile is 'shallower' ($\gamma<4/3$), the
FP diffusion 'pushes' the particles towards the center, making the profile steeper. If the profile
is initially steep, the FP diffusion is directed outwards flatting the cusp. Unfortunately, the
analysis of \citep{evans1997} supposes from the very beginning that the profiles can be only
power-law, and thus this is not a full stability test. Second, as we will see in the next section,
a cusp, very similar to the one under consideration routinely occurs in the halo centers in
cosmological N-body simulations. Though the particle collisions are already significant in this
region \citep{17} (and thus the system may be described by the FP equation), the cusp is formed in
many simulations, being very stable. It would be scarcely possible if $\rho\propto r^{-4/3}$ was an
unstable solution of the FP equation. Such reasoning, however, cannot replace an exhaustive test of
the asymptotical stability of the solution. To summarize: there are strong arguments testifying
that $\rho\propto r^{-4/3}$ is an attractor solution of the FP equation in the case of anisotropic
velocity distribution of particles. However, we cannot irrefutably prove the attractor nature of
the stationary solution, since no full stability test has been performed yet.

\section{Why does the Einasto profile with $n\sim 6$ occur so frequently in N-body simulation?}
The second important conclusion we can draw from Fig.~\ref{18fig1} is that on the main part of the
region under consideration, except the vicinity of the curve $\tau_r/\tau_{FP}=0$, $\tau_{FP}$ is
smaller than $\tau_r$, and even more so, Power's time $1.7 \tau_r$ (see below). It means that the
FP diffusion is, generally speaking, very effective, and the collisions change the profile
parameters in a time shorter than $\tau_r$. It questions the generally accepted criteria of N-body
simulation convergence since all of them in current use are entirely based on merely the density
profile stability. Let us consider, for instance, the most popular criterion $\tau_r\ge 0.6 t_0$
\citep{power2003}. The authors used the method for keeping track of how the averaged density inside
some radius $r$ from the halo center changes with time. They found that a cuspy density profile
(quite similar for various halos) was rapidly formed in the halo center, then the averaged density
inside $r$ remained almost constant until, at least, $t_0\sim 1.7 \tau_r$ at this radius, and then
a core at the halo center appeared. The authors of the criterion \citep{power2003} believed that
the profile universality and stability proved the simulation convergence, and the core formation
was the first sign of the collision influence. The conclusion that the relaxation has no effect
until almost two relaxation times seems rather surprising. However, further tests (conducted with
the use of the same profile stability method) showed that the criterion might be even softer, and
the collision influence might be negligible until tens of relaxation times \citep{hayashi2003,
klypin2013}.

The existence of the stationary solution of the FP equation clearly reveals the vulnerability of
the convergence criteria based on the profile stability: the collisions themselves may form
stationary profiles. The shape of the stationary solution needs not precisely coincide with
$\rho\propto r^{-4/3}$: this profile corresponds to the case when the particle velocity
distribution is isotropic at each point, the halo is spherically symmetric, and the particle
interaction is newtonian or, at least, softened newtonian. All these suppositions are typically not
quite true in real N-body simulations: even the particle interaction is calculated with some
errors, which probably leads to quite observable effects (see below). However, it seems that the
stationary profile in the halo center should lie between $\rho\propto r^{-1}$ \citep{13} and
$\rho\propto r^{-4/3}$ \citep{evans1997}. Figure~\ref{18fig1} shows that any profile that
significantly differs from it is significantly shifted by the collisions in a time $t\lesssim
\tau_r$. On the contrary, the profile corresponding to the stationary solution should be standard
(the FP coefficients are similar for various N-body codes) and very stable, i.e., survive for tens
of relaxation times, as reported in \citep{hayashi2003, klypin2013}. However, it may already be
created by the test body collisions, and thus be no more than a numerical effect. In this case, the
profile universality and stability have nothing to do with the simulation veracity.

However, one may distinguish  such a pseudo-convergence from the real one \citep{13}. The idea is
that the profile stability has different physical reasons in this case. If the collisional
contribution is negligible (as it should be) and the density profile is stable, the energies of
each particle in it remain constant, and thus the particles have stationary orbits. If the
collisions are significant, they lead to the FP diffusion of the particle parameters in the phase
space. Thus, the energies of the particles do not conserve; the particle orbits move up and down
along the cusp. The profile is stable in the case of the stationary solution because the upward and
downward FP streams compensate each other.

An isolated Hernquist halo was simulated in \citep{17}. The Hernquist model is close to the
Navarro-Frenk-White and behaves the same way ($\rho\propto r^{-1}$) in the central region, but it
has the known analytical solution for the stationary velocity distribution. Contrary to standard
N-body simulations, the gravitational potential $\phi(r)$ is exactly constant and spherically
symmetric in this case and, therefore, the energy and angular momentum of each particle should be
conserved. The simulations with the modern version of the Gadget-3 code \citep{springel2005}
confirm that the density profile is stable until $t_0\sim 1.7 \tau_r$, in accordance with
\citep{power2003}. However, it turns out that all integrals of motion characterizing individual
particles experience strong unphysical variations, revealing an effective interaction between them.
Moreover, the authors find Fokker-Planck streams in the cusp region, strong enough to change the
shape of the cusp or even to create it. The cusp is stable, because the upward and downward FP
streams compensate each other, and the stability has nothing to do with the negligibility of
collisional relaxation.

It is probably on account of this fact that the Einasto profile with $n\simeq 6$ occurs so
frequently in N-body modelling. Indeed, the halo profile occurring in the simulations may be
approximated by various phenomenological models at $r\sim r_s$; the Einasto profile with $n\simeq
6$ becomes clearly preferable to the NFW one only at $r\sim 10^{-1} r_s$ (see, for instance, Fig.~3
in \citep{navarro2010}) or, which is approximately equal\footnote{If we quite realistically accept
the average ratio $R_{vir}/r_s\equiv c_{vir}\sim 5$.}, at $r\sim 2\% R_{vir}$ (Fig.~3 in
\citep{dutton2014}). The point $r= 10^{-1} r_s$, $n\simeq 6.2$ lies exactly on the stationary
solution in Fig.~\ref{18fig1}. If the influence of the test particle interaction is already
significant at these radii, we may surmise that it is just the particle collisions that transform
the profile into the Einasto with $n\simeq 6$ here. If the profile is a stationary solution of the
FP equation, it explains both its phenomenal stability up to many relaxation times and
universality: the shape of the profile formed by collisions is defined by the potential of the
particle interaction (which is always close to the newtonian one and therefore universal), and not
by the properties of initial perturbations or by details of the numerical scheme. On the other
hand, probably the profile has nothing to do with the properties of real collisionless dark matter
systems. At least, the profile stability says nothing about the simulation convergence and
reliability.

However, if $\rho\propto r^{-4/3}$ is a stationary solution, why a core finally appears in the
center of the halo? There can be several possible reasons. First, the Fokker-Planck equation is not
exact: it can be obtained from the exact Boltzmann equation by expanding the collision integral in
a Taylor series and dropping all the terms except the first two ones \citep{ll10}. However, the
contribution of the kinetic terms of higher orders also becomes important after several relaxation
times, and it can destroy the cusp (see \citep[eqn. 4, 5]{13} for details). The core formation may
be related to short-range, large-angle deflections, which are not taken into account by the
Fokker-Planck equation. Finally, the core formation may be caused by some other numerical effects.
Whatever the reason is, the core formation cannot be used as the first sign of the collision
influence, as was done in \citep{power2003}: the FP diffusion becomes important much earlier.

Apparently, the profile tends to $\rho\propto r^{-4/3}$ only  for $r\ll r_s$: if $r\sim r_s$ or
larger, the influence of collisions is not strong enough since the relaxation time rapidly grows
with radius.

Two questions arise: what does the model under consideration predict and is the FP diffusion strong
enough to drive the density profiles of halos in numerical simulations towards the stable FP
solution? It seems in the framework of the theory that halos resolved with fewer particles (i.e.,
less massive halos in a cosmological simulation), should have been affected out to larger radii
(since $\tau_r$ is smaller in this case), and therefore follow the Einasto profile with $n\simeq 6$
(hereafter E6 profile) out to larger radii than more massive halos. Furthermore, one should expect
that the central regions of N-body halos have a $\rho\propto r^{-4/3}$ cusp, rather than the
$n\simeq 6$ Einasto profile.

To be specific in answering the questions, let us consider one of the most recent and
high-performance simulations \citep{dutton2014} as an example. Figure~3 in \citep{dutton2014} shows
that the density profile at $(2-4)\% \rvir$ is indeed quite close to E6. The profile is slightly
steeper than E6 for small halos and slightly shallower than E6 for the largest halos. Moreover, the
profile steepness at $3\% \rvir$ regularly decreases with the halo mass growth. Though these
results agree with our theoretical predictions, the statistics is rather poor and does not allow
one to make strong statements, which is quite expectable. Figure~\ref{18fig1} in our paper shows
that in order to reach the region where the stationary solution gets significantly steeper than E6,
one needs to approach closer than $2\% \rvir$ to the halo center. The masses of the $\rho\propto
r^{-1}$ cusp and the $\rho\propto r^{-4/3}$ one differ by $20\%$. Therefore, we need to have at
least $N_m \sim 225$ particles inside $2\% \rvir$ to distinguish the cusps on the $3\sigma$-level
(indeed, $\sigma\simeq \sqrt{N_m}$, and $3\sigma/N_m = 3/\sqrt{225} = 20\%$). For instance, only
$0.46\%$ of the total mass of an NFW halo with $\cvir=5$ lie inside $2\% \rvir$. Thus, if a halo
contains less than $50000$ particles, any profile reconstruction inside $2\% \rvir$ is certainly
impossible.

In fact, this number should be significantly larger in cosmological simulations: the halos formed
there are not exactly spherical and experience tidal perturbations; their main structural
parameters are significantly scattered (the halos form a random gaussian field, see
\citep[Fig.~15]{dutton2014}) and in addition to that are measured with an accuracy of $(15-25)\%$,
depending on the algorithm used. The apparently numerical core formation in the halo center is also
a serious handicap to the profile reconstruction. As a result of all these factors, the authors of
\citep{dutton2014} report that $\sim 10^4$ test particles are necessary just to obtain reliable
$r_s$ and other Einasto shape parameters. We can conclude that a halo should contain hundreds of
thousand particles to distinguish $\rho\propto r^{-1}$ and $\rho\propto r^{-4/3}$ cusps inside $2\%
\rvir$ in real cosmological simulations.

A set of simulations with various particle masses was used in \citep{dutton2014} in order to
increase the halo mass range covered by the simulation and the result reliability (see the paper
for details). A single simulation contains $\sim 10^4$ halos holding more than $500$ particles and
no more than $(600)^3\simeq 2\cdot 10^8$ particles in total. Unfortunately, the authors have not
specified the mass distribution of the halos. However, we can expect that it obeys the usual law
$d\eta\propto M^{-2} dM$ \citep{diemand2005}, which implies that each next logarithmic mass
interval $[10^{i+1}M_\odot; 10^{i+2}M_\odot]$ contains $10$ times less halos than the previous one
$[10^{i}M_\odot; 10^{i+1}M_\odot]$. Then among $\sim 10^4$ halos containing $\ge 500$ particles
there should be $\sim 10$ containing $\ge 5\cdot 10^5$ ones. Thus, even recent and high-performance
simulations contain only tens of halos, for which the profile reconstruction below $2\% \rvir$ is
possible. Considering the random gaussian scattering of their properties and other difficulties, it
makes it possible to find the above-mentioned tendencies that confirm the predictions of our
theory, but the cosmological simulations should contain, at least, $\sim 10$ times more test
particles in order to make statements more certain.

Strong influence of the FP diffusion in N-body simulations has been demonstrated in a conclusive
way by modelling an isolated halo \citep{17}. In contrast with cosmological simulations, the halo
in this case is spherically symmetric and does not experience tidal effects. If we properly set the
initial velocities, not only the density profile but the particle velocity distribution in each
point and even the energy and the vector of angular momentum of each particle should be conserved.
Any deviation from this behavior is a numerical effect. Thus, one may collect much more information
about the N-body scheme properties than in the case of cosmological simulations. Simulations
\citep{17} of an isolated Hernquist halo revealed that the numerical FP streams are strong enough
to create the cusps. Recent simulations \citep{bosch2018} show that some numerical effect
stabilizes the central part of subhalos and prevents their tidal destruction in N-body simulations.
Indeed, the unphysical cusp formation may be the factor that makes the small halos more robust
\citep{19}. Finally, our analysis also confirms that the FP diffusion is strong enough to influence
the halo profile.

Can we state from our analysis that the FP diffusion in N-body simulations tends to transform the
profile in the halo center exactly into $\rho\propto r^{-4/3}$? Certainly not. First, the
assumptions applied to deduce the stationary solution in this paper or \citep{evans1997} can fail
in the case of N-body simulations, in particular, velocity distribution of the particles need not
be isotropic at each point.

A more fundamental problem is that the above-mentioned papers \citep{bosch2018} and \citep{17}
report about significant effects in the halo center, but their origin has been clarified. The
numerical effects may have be caused not only by the weak encounters, but also by inaccuracies of
the potential calculation, for instance. Equations (\ref{18a8}) for the FP coefficients were
obtained on the assumption that the particle interaction is newtonian, or, at least, truncated
newtonian. If it is not exactly so, the coefficients may differ from (\ref{18a8}).

However, the fact that the universal density profile obtained in simulations falls exactly on the
stationary solution of the Fokker-Planck equation in the halo center does not look like a mere
coincidence and casts doubt on the convention convergence criteria of N-body simulations. The
criteria based on the profile stability are apparently insufficient, and new methods should be
developed and used. Some techniques were offered in \citep{13}. A better understanding of the
possible influence of the inaccuracies of the potential calculation algorithm used in N-body codes
is necessary for a reliable interpretation of simulation results.

\section{Acknowledgements}
We acknowledge the support by the Helmholtz Alliance for Astroparticle Physics HAP funded by the
Initiative and Networking Fund of the Helmholtz Association.  B.M.V. acknowledges support from the
Japan Society for the Promotion of Science (Grant no. 16H00878), NSF  grant AST-1306672 and DoE
grant DE-SC0016369.


\end{document}